\documentclass[11pt]{article}

\usepackage{ifpdf}
\usepackage{amsmath}
\usepackage{amssymb}
\usepackage{amsthm}
\usepackage{color}
\usepackage{fullpage}

\ifpdf
    \usepackage[pdftex]{graphicx}
    \usepackage{epsfig}
\else
    \usepackage{graphicx}
    \usepackage{epic,eepic}
\fi

\newtheorem{theorem}{Theorem}
\newtheorem{lemma}[theorem]{Lemma}
\newtheorem{prop}[theorem]{Proposition}
\newtheorem{cor}[theorem]{Corollary}
\newtheorem{obs}[theorem]{Observation}
\theoremstyle{definition}
\newtheorem{defn}[theorem]{Definition}

\newtheorem*{problem}{Problem}

\newtheorem{open-problem}[theorem]{Open Problem}

\newcommand{\abs}[1]{\left\lvert#1\right\rvert}
\newcommand{\class}[1]{\textup{\textbf{#1}}}
\newcommand{\prob}[1]{\textup{\textsc{#1}}}

\newcommand{\ceil}[1]{\left\lceil#1\right\rceil}
\newcommand{\floor}[1]{\left\lfloor#1\right\rfloor}

\newenvironment{namedtheorem}[1]
           {\begin{trivlist}
         \item {\bf #1.}\em}
           {\end{trivlist}}

\begin{document}

\title{The overlap number of a graph}

\author{Bill Rosgen\\
    Centre for Quantum Technologies\\
    National University of Singapore\\
    bill.rosgen@nus.edu.sg
    \and
    Lorna Stewart\\
    Department of Computing Science\\
    University of Alberta\\
    Edmonton, Alberta, Canada\\
    stewart@cs.ualberta.ca}

\maketitle

\begin{abstract}
  An overlap representation is an assignment of sets to the vertices
  of a graph in such a way that two vertices are adjacent if and only
  if the sets assigned to them overlap.  The overlap number of a graph
  is the minimum number of elements needed to form such a
  representation.  We find the overlap numbers of cliques and complete
  bipartite graphs by relating the problem to previous research in
  combinatorics.  The overlap numbers of paths, cycles, and
  caterpillars are also established.
  Finally, we show the \class{NP}-completeness of the problems of
  extending an overlap representation and finding a minimum overlap
  representation with limited containment.
 \end{abstract}
 
\section{Introduction}
\label{scn:introduction}

All graphs we consider are finite and simple.  
The subgraph of $G = (V,E)$ induced by $V' \subseteq V$ is denoted by
$G[V']$.
For graph $G=(V,E)$ and vertex $v \in V$, $N(v) = \{ u~|~(u,v) \in E \}$ is called the {\em open neighbourhood} of vertex $v$;
$N[v] = N(v) \cup \{v\}$ is the {\em closed neighbourhood} of vertex $v$.
The minimum degree of any vertex in a graph $G$ is denoted $\delta(G)$.
A set of vertices is a \emph{clique} 
if any two vertices in the set are adjacent.  
A clique
is \emph{maximal} if it is not contained in a larger clique.  An
\emph{independent set} is a set of pairwise nonadjacent vertices.  We
denote by $K_n$ the complete graph on $n$ vertices.  We use
the notation $P_n$, for the path on $n$ vertices, and $C_n$ for
the cycle on $n$ vertices.  We also need to introduce two covers
of a graph: a \emph{clique cover} is a covering of the vertices of a
graph by cliques, and an \emph{edge-clique cover} is a covering of the
edges of a graph by cliques.

Two sets \emph{overlap} if they intersect and neither set contains the
other.  An \emph{overlap representation} (respectively
\emph{intersection representation}) for a graph $G$ is an assignment
of sets to the vertices of $G$ such that two vertices are adjacent if
and only if the sets assigned to them overlap (respectively
intersect).  The size of such a representation is the cardinality of
the union of the assigned sets, and the minimum size of a
representation is termed the \emph{overlap number} (respectively
\emph{intersection number}) of the graph.  
The overlap number of graph $G$ is denoted $\varphi(G)$.
Every graph has an overlap representation.  This follows from the fact
that all graphs have intersection
representations~\cite{marczewski45sur} (for an English translation, see \cite{translation}), and the observation that we can take an
intersection representation for a graph and add a new element to each
set, which causes sets to overlap if and only if they
intersect.
While intersection representations of graphs have been widely studied,
overlap representations
have received considerably less attention, even though overlapping is a natural relation of
pairs of sets.

The intersection number parameter was introduced and bounded by
Erd{\H o}s, Goodman, and P{\' o}sa~\cite{egp66representation}.
Specifically, they give a tight upper bound of
$\floor{n^2/4}$ on the intersection number of an $n$-vertex graph.
The
\class{NP}-completeness of computing the intersection number is shown
by Kou, Stockmeyer, and Wong~\cite{ksw78covering}.  In addition to
these results, Raychaudhuri gives polynomial time algorithms for the
intersection numbers of chordal
graphs~\cite{raychaudhuri88intersection} and $W_4$-free
comparability graphs~\cite{raychaudhuri91edge}, where $W_4$ is a cycle
on four vertices with a universal vertex.  

For the overlap number, only a few algorithms and bounds are known.
Using the simple technique of adding
a new vertex to each set in the representation, the intersection number bound of~\cite{egp66representation} shows
that $\varphi(G) \leq \floor{n^2/4} +n$ for any graph $G$ with $n$
vertices.  
It is known that any cocomparability graph $G$ on $n$ vertices has $\varphi(G) \le
n+1$, since a containment representation of size $n+1$ exists for
$\overline{G}$ in which every set has a common element~\cite{gs89containment}, 
and such a representation is an overlap representation for $G$. Thus
$K_{\frac{n}{2},\frac{n}{2}}$ has intersection number $\floor{n^2/4}$
\cite{egp66representation} and overlap number at most $n+1$ since it is a
cocomparability graph.
Any graph with at most a linear number of
maximal cliques must have linear intersection number, as a
minimum intersection representation and a minimum edge-clique cover
have the same size~\cite{egp66representation}.  
This implies that a graph with a linear number of maximal cliques
must also have linear overlap number, since we can add a new element to
each set of a minimum intersection representation.  This technique yields
linear bounds on the overlap number of trees, chordal graphs, and
planar graphs, using previous linear bounds on the number of maximal
cliques in chordal graphs~\cite{fg65incidence} and planar
graphs~\cite{prisner92few}.
Henderson \cite{henderson} gives lower bounds on the overlap number 
of a graph in terms of its independent sets,
and 
constant factor
approximation algorithms for the overlap numbers of trees and planar graphs. In addition, he shows that there exist graphs with overlap number quadratic in the number of vertices of the graph. Cranston et al \cite{west} show how to compute the overlap number of a tree in linear time, and give upper bounds on the overlap numbers of some graphs. Their results include
the following bounds which are satisfied with equality for some graphs:
$\varphi(G) \leq |E|-1$ where $G=(V,E)$, $G \ne K_3$, and $\delta(G) \ge 2$,
and
$\varphi(G) \leq n^2/4 -n/2 -1$ where $G$ is an $n$-vertex graph with $n \ge 14$.

In this paper we present hardness
results for problems related to finding the overlap number, and give formulas 
and describe algorithms for
the overlap numbers of some simple graphs.
These results appeared in the first author's Master's thesis~\cite{rosgen}.

The remainder of this paper is organized as follows.  In
Section~\ref{scn:overlap} we formally introduce the overlap number and
discuss some of the basic properties of overlap representations.  
We follow this with Section~\ref{scn:algo}, where we give formulas and describe algorithms
for the overlap numbers of some graphs.
Finally, in Section~\ref{scn:hardness}, we present some
\class{NP}-completeness results on problems related to the overlap number.

\section{The Overlap Number of a Graph}
\label{scn:overlap}

Before formalizing the definition of an overlap representation, we
note that by a \emph{collection} we refer to a multiset of sets, and for
simplicity we allow the mapping from sets of a representation to
vertices of a graph to remain implicit, with the set $S_v$ associated
with the vertex $v$ of the graph.  With these notational issues
resolved, we define an overlap representation in the following way.

\begin{defn}\label{defn:overlap}
  Given a graph $G = (V,E)$, a collection $\mathcal{C} = \{S_v : v \in
  V\}$ is an \emph{overlap representation} for $G$ if for every $u,v \in V$
  we have
  \[ (u,v) \in E \text{ if and only if } S_u \cap S_v \neq \emptyset,
     S_u \not\subseteq S_v, \text{ and } S_v \not\subseteq S_u. \]
  We define the \emph{size} of a representation to be the number of 
  elements used in the representation, which is
  $ \abs{ \bigcup_{v \in V} S_v }, $ and we let the \emph{overlap
  number}, $\varphi(G)$, be the size of a minimum overlap
  representation.
\end{defn}

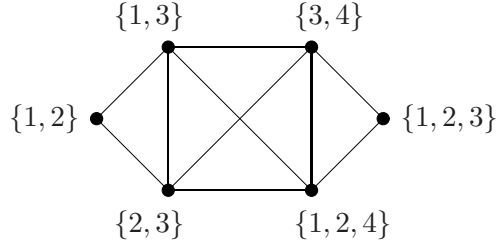
\begin{figure}
  \begin{center}
    \setlength{\unitlength}{3947sp}%
\begingroup\makeatletter\ifx\SetFigFont\undefined%
\gdef\SetFigFont#1#2#3#4#5{%
  \reset@font\fontsize{#1}{#2pt}%
  \fontfamily{#3}\fontseries{#4}\fontshape{#5}%
  \selectfont}%
\fi\endgroup%
\begin{picture}(2953,1493)(701,-1570)
{\color[rgb]{0,0,0}\thinlines
\put(2551,-1261){\circle*{76}}
}%
{\color[rgb]{0,0,0}\put(2551,-361){\circle*{76}}
}%
{\color[rgb]{0,0,0}\put(3001,-811){\circle*{76}}
}%
{\color[rgb]{0,0,0}\put(1201,-811){\circle*{76}}
}%
{\color[rgb]{0,0,0}\put(1651,-361){\circle*{76}}
}%
{\color[rgb]{0,0,0}\put(1651,-1261){\line( 0, 1){900}}
}%
{\color[rgb]{0,0,0}\put(1651,-361){\line( 1, 0){900}}
}%
{\color[rgb]{0,0,0}\put(2551,-361){\line( 0,-1){900}}
}%
{\color[rgb]{0,0,0}\put(2551,-1261){\line(-1, 0){900}}
}%
{\color[rgb]{0,0,0}\put(1651,-1261){\line(-1, 1){450}}
}%
{\color[rgb]{0,0,0}\put(1201,-811){\line( 1, 1){450}}
}%
{\color[rgb]{0,0,0}\put(2551,-1261){\line( 1, 1){450}}
}%
{\color[rgb]{0,0,0}\put(3001,-811){\line(-1, 1){450}}
}%
{\color[rgb]{0,0,0}\put(1651,-1261){\line( 1, 1){900}}
}%
{\color[rgb]{0,0,0}\put(2551,-1261){\line(-1, 1){900}}
}%
\put(651,-901){\makebox(0,0)[lb]{$\{1,2\}$}}%
\put(1311,-261){\makebox(0,0)[lb]{$\{1,3\}$}}%
\put(1311,-1566){\makebox(0,0)[lb]{$\{2,3\}$}}%
\put(2441,-261){\makebox(0,0)[lb]{$\{3,4\}$}}%
\put(3111,-901){\makebox(0,0)[lb]{$\{1,2,3\}$}}%
\put(2441,-1566){\makebox(0,0)[lb]{$\{1,2,4\}$}}%
{\color[rgb]{0,0,0}\put(1651,-1261){\circle*{76}}
}%
\end{picture}%
  \end{center}
  \caption[Example of a minimum overlap representation]{Example of a
    minimum overlap representation.}
  \label{fig:representation}
\end{figure}

As can be seen from the definition, overlap representations have many
similarities to intersection representations: sets assigned to
adjacent vertices must intersect and disjoint sets map to nonadjacent
vertices.  In the overlap case, however, the situation is
more complex, as not only do we need to ensure a stronger condition
than intersection for adjacent vertices, we have a choice of
representation, for every non-edge, of disjointedness or containment.
As an example, consider the representation in
Figure~\ref{fig:representation}, where there are nonadjacent
vertices represented in each of these ways.

In the case of an
intersection representation, if we take an element of the
representation and examine all those sets it is contained in, we find
that the vertices associated with them form a clique.  Doing the same
in an overlap representation once again leaves us with a collection of
vertices with intersecting sets, except here we may have non-edges
represented by containment, and so, since the orientation implied by
set containment forms a partial order, we can map elements of the
representation to cocomparability graphs.  Unfortunately, while
covering all edges of a graph with cliques leads to an intersection
representation, if we simply cover the edges of a graph with
cocomparability graphs, we do not generally end up with a valid
overlap representation (most cocomparability graphs do
not have overlap number one).  

\begin{obs}\label{obs:subgraph}
  If $\{S_v : v \in V \}$ is an overlap representation for $G=(V,E)$
  then, for any $V' \subseteq V$, $\{S_v : v \in V' \}$ is an overlap
  representation for $G[V']$.  Thus $\varphi(G) \ge \varphi(H)$ for all
  induced subgraphs $H$ of $G$.
\end{obs}

\emph{Vertex multiplication} is the expansion of a vertex into an independent
set, such that the vertices of the independent set have the same
adjacencies as the original vertex.

\begin{obs}\label{obs:multiplication}
  If there is an overlap representation of size $s$ for graph $G$ then
  there is an overlap representation of size $s$ for graph $G'$, where
  $G'$ arises from $G$ by vertex multiplication.
\end{obs}

This can be observed by simply duplicating the set assigned to a
vertex when it is multiplied.  Note that the size of an intersection
representation is preserved by the operation of expanding a vertex to
a clique but not by vertex multiplication.

We conclude this section with three results that will be used in the next section.

\begin{lemma}\label{lemma:walk}
  If $A$, $B$, and $C$ are three sets such that $A \subseteq C$, $A$
  overlaps $B$, but $B$ does not overlap $C$, then $B \subseteq C$.

  \begin{proof}
    Since $A$ and $B$ overlap, we have $A \cap B \ne \emptyset$ and $A \backslash B \ne \emptyset$ which, combined with the fact that $A \subseteq C$, imply 
    $C \cap B \ne \emptyset$ and $C \backslash B \ne \emptyset$.
    Now since $B$ does not overlap $C$, $B \subseteq C$. 
     \end{proof}
\end{lemma}

We amplify this lemma to the following stronger result that we
use in Section~\ref{scn:components}
to argue bounds on the size
of an overlap representation for a graph, based on the size of
representations for the connected components of the graph.

\begin{lemma}\label{lemma:basic-contain}
  Let $G = (V,E)$ be a graph
  with overlap representation
  $\mathcal{C} = \{S_v : v \in V\}$.  
  Let $X$ and $Y$ be nonempty subsets of $V$ such that each of $G[X]$ and
  $G[Y]$ is connected, and
  no edge of $E$ has one endpoint in $X$ and the other in $Y$.
  Let
  $U_X = \bigcup_{x \in X} S_x$
  and $U_Y = \bigcup_{y \in Y} S_y$.
  If $S_x \subseteq S_y$ for some $x \in X$, $y \in Y$, then
  \begin{description}
  \item[(i)]
    for all $y \in Y$, 
    $U_X \subseteq S_y$ or $U_X \cap S_y = \emptyset$, 
  and 
  \item[(ii)]
  if $|X|>1$ or $|Y|>1$ then
  for all $x \in X$, $y \in Y$, 
    $S_y \not\subseteq S_x$.
  \end{description}
  
  \begin{proof}
  We first show that, for all $x \in X, y \in Y$,
  
\begin{center}  
(1) if $S_x \subseteq S_y$ then $U_X \subseteq S_y$,
(2) if $S_y \subseteq S_x$ then $U_Y \subseteq S_x$,
and
(3) if $S_x= S_y$ then $|X|=|Y|=1$.
\end{center}

Suppose (1) is false.
Let $x' \in X$ be such that $S_{x'}$ contains an element that is not in $S_y$,
where the distance in $G[X]$ from $x$ to $x'$ is as small as possible.
Let $x = x_1 x_2 \ldots x_k = x'$ be a shortest $x, x'$-path in $G[X]$.
Then $S_{x_{k-1}} \subseteq S_y$ (by the choice of $x'$),
$S_{x_{k-1}}$ overlaps $S_{x'}$ (because $x_{k-1}$ and $x'$ are adjacent on the path), 
and $S_{x'}$ does not overlap $S_y$ (since $x'$ and $y$ are not adjacent in $G$).
But then by Lemma \ref{lemma:walk}, $S_{x'} \subseteq S_y$, contradicting the choice of $x'$.
The justification for (2) is similar.
For (3), if $S_x= S_y$ then $x$ and $y$ are adjacent to exactly the same vertices in $G$,
and therefore, since each of $G[X]$ and $G[Y]$
is connected, and there are no edges between $X$ and $Y$ in $G$,
$|X|=|Y|=1$.

Note that (i) is true if $|X|=|Y|=1$ since then $U_X = S_x \subseteq S_y$ for $x \in X$, $y \in Y$.
Now suppose (i) is false.
Then 
$|X|>1$ or $|Y|>1$ and
there exists $y \in Y$ with $U_X \not\subseteq S_y$ and $U_X \cap S_y \ne \emptyset$.
By (1), there is no $x \in X$ with $S_x \subseteq S_y$. 
Therefore, $S_y \subseteq S_{x'}$
for some $x' \in X$.
We also have $S_x \subseteq S_{y'}$ for some $x \in X$, $y' \in Y$, by the statement of the lemma.
But now, by (1) and (2), $U_X \subseteq S_{y'} \subseteq U_Y \subseteq S_{x'} \subseteq U_X$,
which implies $S_{x'} = S_{y'} $, contradicting (3).

If (ii) is false, we again have $S_x \subseteq S_{y'}$ and $S_y \subseteq S_{x'}$
for some $x,x' \in X$ and $y,y' \in Y$ which, together with (1) and (2), contradicts (3).
  \end{proof}
\end{lemma}

In Section~\ref{scn:paths} we use the following simplification of
Lemma \ref{lemma:basic-contain} to argue bounds on the overlap numbers of paths, cycles,
and caterpillars.

\begin{cor}\label{cor:contain-connected}
  Let $G = (V,E)$ be a graph, and let $\mathcal{C} = \{S_v : v \in
  V\}$ be an overlap representation of $G$.  Fix $v \in V$, and let,
  for $u \in V \setminus N[v]$, $A_v(u)$ be the vertex set of the
  connected component of $G[V \setminus N[v]]$ that contains $u$.  If
  $S_u \subseteq S_v$, then
  $ \bigcup_{w \in A_v(u)} S_w \subseteq S_v. $
\end{cor}

\section{Minimum Overlap Representations}
\label{scn:algo}

In this section, we give formulas for the overlap numbers of cliques, complete $k$-partite graphs, paths, cycles,
and caterpillars and for the overlap
number of a disconnected graph in terms of the overlap numbers of its
connected components.

\subsection{Cliques and Complete $k$-partite Graphs}\label{scn:clique}

An overlap representation of a clique is simply a collection of sets
where no set contains any other and no two sets are disjoint. We can
apply a theorem of Milner to find the minimum size of such a
representation.

\begin{defn}\label{defn:int_anti}
  The maximum size of a family, $\mathcal{C}$, of subsets of
  $\{1,2, \ldots, m\}$ satisfying, for $p \geq 0$,
  \begin{enumerate}
    \item If $A,B \in \mathcal{C}$, with $A \neq B$, then $A \not\subseteq B$,
    \item If $A,B \in \mathcal{C}$, then $\abs{A \cap B} \geq p$,
  \end{enumerate}
  is denoted $S(p,m)$.
\end{defn}
  
The value of the function $S(p,m)$ is exactly the quantity given by
Milner's Theorem, first published in 1966.

\begin{theorem}[Milner \cite{milner66combinatorial}]
  \label{thm:milner}
  For $m \geq 1$ and $p \geq 0$,
  \[ S(p,m) = \binom{m}{\floor{\frac{m+p+1}{2}}}. \]
\end{theorem}

Also noted by Milner~\cite{milner66combinatorial}, is that it is easy
to construct a collection that achieves this bound, by simply choosing
all subsets of $\{1,2, \ldots, m\}$ of size $\floor{(m+p+1)/2}$.
This, reformulated in the language of overlap representations, is
precisely the content of the following corollary.

\begin{cor}\label{cor:min-p-overlap-clique}
For $n \ge 1$, 
$\varphi(K_n) = \min \left\{m : n \leq S(1,m) \right\}$.

  \begin{proof}
    Consider any overlap representation, $\mathcal{B}$, of $K_n$.
    Any two elements of $\mathcal{B}$ must intersect,
    and no element can contain any other, as each pair
    of vertices in $K_n$ forms an edge.  Thus
    $\abs{\mathcal{B}} \leq S(1,m)$, where $m = \abs{\bigcup_{A \in
    \mathcal{B}} A}$.

    For any $m$, consider the collection given by $\mathcal{C}_m = \{A
    \subseteq \{1,2, \ldots, m\} : \abs{A} = \floor{(m+2)/2}\}$.  As we
    have $\floor{(m+2)/2} \geq \ceil{m/2}$, any two elements of
    $\mathcal{C}_m$ form an intersecting pair, and furthermore, no
    element is contained in any other, as they all have the same size.
    Counting the number of ways to form subsets of $\{1, 2, \ldots,
    m\}$, we obtain
    \[ \abs{\mathcal{C}_m} = \binom{m}{\floor{\frac{m+2}{2}}}
                           = S(1,m). \]

    Then, to find the minimum representation, we seek the minimum $m$
    that leaves enough room to form an overlap representation.  We can
    simply choose any $n$ elements of $\mathcal{C}_m$ to obtain an
    overlap representation on $m$ elements.  Thus, $\varphi(K_n)$ is the
    smallest $m$ such that $ n \leq \abs{\mathcal{C}_m} = S(1,m), $ as
    desired.
  \end{proof}
\end{cor}
				   
The next result follows immediately from
Corollary~\ref{cor:min-p-overlap-clique} and
Observation~\ref{obs:multiplication}.

\begin{cor}\label{cor:k-partite}
  If $G$ is a complete $k$-partite graph, 
  then $\varphi(G) = \min \{m : k \leq S(1,m)\}$.
\end{cor}

We now investigate some computational issues involved in finding
overlap representations of cliques and $k$-partite graphs. This is
done by first finding bounds on $\varphi(K_n)$ in terms of $n$ which,
together with the constructive proof of
Corollary~\ref{cor:min-p-overlap-clique}, yield a simple polynomial
time algorithm to produce a minimum overlap representation of $K_n$.

In order to gain a view of the size of the required
representation for a given graph, we unwind the expression
\[ \min \left\{m : n \leq \binom{m}{\floor{\frac{m+2}{2}}} \right\}, \]
to obtain an asymptotically tight bound on $\varphi(K_n)$ in terms of $n$.
We make use of Stirling's Approximation, which can
be found, for example, in \cite{bollobas98modern}:
\begin{equation}\label{eqn:stirling}
  \sqrt{2 \pi n} (n/e)^n \leq n! \leq e^{1/(12n)} \sqrt{2 \pi n} (n/e)^n.
\end{equation}
This results in the following lemma.

\begin{lemma}\label{lemma:messy_stirling}
  For $1 \leq k < n$,
  \[ \binom{n}{k} \geq \sqrt{\frac{1}{8 \pi k}}
      \left(\frac{n}{k}\right)^k \left(\frac{n}{n-k}\right)^{n-k}. \]
  \begin{proof}
    The inequality follows by substituting
    equation~\eqref{eqn:stirling} into the expansion of the binomial
    coefficient.
  \end{proof}
\end{lemma}

Using this lemma, we bound the size of the minimum overlap
representation of the graphs we have considered.  The proof here is
simply a calculation and is omitted.

\begin{theorem}\label{thm:clique-bound}
  For $n \ge 1$,
  \[ \min \left\{m : n 
    \leq \binom{m}{\floor{\frac{m+2}{2}}} \right\} \in \Theta(\log n).\] 
\end{theorem}

The next result is the final ingredient needed to build an
efficient algorithm to find, for a given $n$, the minimum $m$ such
that $n \leq S(1,m)$.

\begin{prop}\label{prop:recurrence}
  For any $m \geq 2$,
  \[ S(1,m) = 
     \begin{cases} 
       \frac{2m}{m-1} S(1,m-1) & \text{if $m$ is odd}, \\
       \frac{2m}{m+2} S(1,m-1) & \text{if $m$ is even}.
     \end{cases} \]

  \begin{proof}
    The proof makes use of the following identities on binomial coefficients
    \begin{eqnarray*}
      \binom{n}{k} &=& \frac{n}{n-k}\binom{n-1}{k} \\
      \binom{n}{k} &=& \frac{n}{k}\binom{n-1}{k-1}, 
    \end{eqnarray*}
    which can be found, for example, in~\cite{knuth97volume1}.
  \end{proof}
\end{prop}

Using the recurrence of Proposition~\ref{prop:recurrence}, we can
compute $S(1,m)$ for successive values of $m$, until we find
$\varphi(K_n)$, the smallest $m$ such that $n \leq S(1,m)$.
By Theorem~\ref{thm:clique-bound},
this produces an algorithm with runtime in $O(\log n)$.
To compute a
minimum representation for $K_n$, as noted in the proof of
Corollary~\ref{cor:min-p-overlap-clique}, we simply take any $n$
of the subsets of $\{1,2,\ldots, \varphi(K_n)\}$ of cardinality $\floor{(\varphi(K_n)+2)/2}$.
Since $\varphi(K_n) \in O(\log n)$
(by Theorem~\ref{thm:milner}, Corollary~\ref{cor:min-p-overlap-clique} and Theorem~\ref{thm:clique-bound}),
$n$ of these subsets can be found in $O(n)$
time.
These algorithms can also be immediately extended to find
representations for complete $n$-partite graphs, as described in
Observation~\ref{obs:multiplication}.

\subsection{Paths, Cycles, and Caterpillars}
\label{scn:paths}

A minimum intersection
representation for a path is simple to find, as there is only one possible
edge-clique cover, the one consisting of each maximal clique.
While it is essentially no harder to
find an overlap representation of a path, proving the optimality of
the representation is more difficult.  Once we have shown the size
that an overlap representation of a path must have,
we extend the result to the case of cycles and caterpillars. The
construction, given as part of the proof of the following theorem can
be immediately transformed
into a
efficient algorithm for generating an overlap representation of
$P_n$.  This theorem does not hold for $P_2$, as $\varphi(P_2) = 3$.

\begin{theorem}\label{thm:path}
  For $n \geq 3$, $\varphi(P_n) = n$.
  
  \begin{proof}
    For $n = 3$, we observe that $\{ \{1,2\}, \{2,3\}, \{1,2\} \}$ is
    a minimum overlap representation, since we need at least three
    elements to represent a single edge.     
    We now show that, for $n \ge 4$, $\varphi(P_n) \geq
    \varphi(P_{n-1}) + 1$, thereby proving that $\varphi(P_n) \geq n$.
    For $n \geq 4$, let
    $1,2,\ldots, n$ be the vertices of $P_n$ in the order in which they appear on the path, and 
    let $\mathcal{C} = \{S_1, S_2, \ldots, S_n\}$ be a
    minimum overlap representation for $P_n$.
    By Corollary~\ref{cor:contain-connected}, either $S_1$ contains none
    of $\{S_3, S_4, \ldots, S_n\}$, or it contains each $S_i$ for $i
    \geq 3$.
    Notice also that these
    two cases collapse, since if $S_1$ contains all $S_i$ for $i \geq 3$,
    then in particular, $S_n \subseteq S_1$, and so, if we consider
    the reversal of the path, we find that $S_n$ contains none of the
    other sets, since $n \geq 4$.  We thus need only consider the
    first case.

    To this end, let the representation, without loss of generality,
    be such that the set $S_1$ contains none of $\{S_3, \ldots S_n \}$.
    Notice that with the exception of $S_2$, the elements of $S_1$ are
    either all contained in one of the other sets, or none of them
    are.  We form a representation for $P_{n-1}$ where these
    elements are compressed into a single element.  We consider the
    collection given by $\mathcal{C}' = \{S_2 \cup S_1, S_3, \ldots S_n \}$.
    As $S_1$ and $S_2$ share at least one element, $S_k$ does not
    overlap $S_2 \cup S_1$ for any $k \geq 4$.
    To see this, we consider two cases.  The first case is
    that $S_1 \subseteq S_k$, but then, by
    Lemma~\ref{lemma:walk}, we have $S_2 \subseteq S_k$ as
    well, which implies that $S_1 \cup S_2 \subseteq S_k$, as
    desired.  In the other case we have $S_1$ disjoint from $S_k$, but
    in this case we can observe that $S_2 \not\subseteq S_k$, as this
    would imply, again by Lemma~\ref{lemma:walk}, that $S_1
    \subseteq S_k$.  Since $S_2 \not\subseteq S_k$, it is either
    disjoint from $S_k$, in which case $S_1 \cup S_2$ is as well, or
    $S_k \subseteq S_2$, which implies that $S_k \subseteq S_1 \cup
    S_2$, as required.
    Similarly, in the collection $\mathcal{C''} = \{S_2 \setminus S_1,
    S_3, \ldots, S_n \}$, a case analysis shows that the set 
    $S_2 \setminus S_1$ does not overlap any set $S_k$ for $k \geq 4$.

    Thus, we need only verify that one of these two
    collections preserves the overlap between $S_3$ and the
    replacement for $S_2$.  To see that at least one suffices, let
    $\mathcal{C}'$ fail to be an overlap representation for $P_{n-1}$,
    which implies that $S_3 \subseteq S_2 \cup S_1$, as we have only
    enlarged $S_2$.  Thus, $S_1 \cap S_3 \neq \emptyset$,
    as $S_3$ is contained in neither $S_1$ or $S_2$, but it is
    contained in their union.  Also, since $S_1$ does not contain any
    other set in the representation, we must have $S_1 \subseteq S_3$
    as these two vertices are not adjacent in the path.  Notice also
    that, since $S_3$ is contained in $S_1 \cup S_2$, and $S_3$ is not
    contained in $S_1$, we must have $S_3 \cap (S_2 \setminus S_1)
    \neq \emptyset$.  Seeking a contradiction, we assume that $S_3$
    and $S_2 \setminus S_1$ also do not overlap.  Since $\mathcal{C}$
    is an overlap representation for $P_n$, we must have $S_3
    \not\subseteq S_2 \setminus S_1 \subseteq S_2$, as the vertices
    associated with $S_2$ and $S_3$ are adjacent.  This leaves only
    one way for $S_3$ to fail to overlap $S_2 \setminus S_1$, which is
    $(S_2 \setminus S_1) \subseteq S_3$.  If this is the case, then we
    have $S_2 \subseteq S_3$, as we know that $S_1 \subseteq S_3$,
    which we derived from the failure of $\mathcal{C}'$.  This
    contradicts the fact that $\mathcal{C}$ is an overlap
    representation for $P_n$, and so one of $\mathcal{C}'$ and
    $\mathcal{C}''$ must form a valid representation for $P_{n-1}$.
    In both of these
    representations, each set either contains $S_1$ or is disjoint
    from it, and so there is no loss in replacing the elements of
    $S_1$ with a single element.  This reduces the size of the
    representation by at least one, as a set needs at least two
    elements to overlap another set.  Hence, we have formed a
    representation of $P_{n-1}$ of size at most $\varphi(P_n) - 1$.
    By induction on $n$, we have shown that
    \[ \varphi(P_{n})
       \geq 1 + \varphi(P_{n-1}) 
       = 1 + n - 1 = n. \]
      
    To finish the proof, it is sufficient to build a representation of
    this size.  Consider the representation
    for $P_n$ given by, for $1 \leq i \leq n-1$,
    \begin{eqnarray*}
      S_i & = & \{i,i+1\} \\
      S_n & = & \{1,2,\ldots,n-1\}.
    \end{eqnarray*}
    Notice that in this representation, on the first $n-1$ vertices,
    the set $S_i$ overlaps only the sets $S_{i-1}$ and $S_{i+1}$ and
    is disjoint from the other sets, with the exception of $S_n$.
    Also, $S_n$ contains all sets except $S_{n-1}$, which it overlaps,
    and so this is an overlap representation for $P_n$ using $n$
    elements.
    This proves that $\varphi(P_n) = n$.      
  \end{proof}
\end{theorem}

The representation used in the proof of the theorem is optimal
in the number of elements used, and can be constructed in $O(n)$ time,
which is asymptotically optimal, as a representation needs to have
linear size.  Thus we can view this construction as an efficient
algorithm to find an overlap representation of a path.

Having found the overlap number of a path, we can find immediate lower
bounds on the size of the overlap representation for some other simple
graphs.  The first of these is $C_n$, the cycle on $n$ vertices.  Once
again, the lower bound is matched by a simple construction, which
can be transformed immediately into an algorithm with running time
linear in $n$.  This result is not true for $n=3$, as $\varphi(C_3) =
3$.

\begin{cor}
  For $n \geq 4$, $\varphi(C_n) = n-1$.

  \begin{proof}
    To see that $\varphi(C_n) \geq n-1$ we simply observe that by
    Theorem~\ref{thm:path}, the size of the representation for any
    $n-1$ of the $n$ vertices is at least $n-1$, and so it remains
    only to construct a representation using $n-1$ elements.  We do
    this by setting, for $1 \leq i \leq n-2$,
    $ S_i = \{i,i+1\}, $
    which forms an overlap representation for a path of $n-2$ vertices,
    using $n-1$ elements.  We add to this representation $S_{n-1} =
    \{1,2,3,\ldots,n-2\}$ and $S_n = \{2,3,4,\ldots,,n-1\}$, noting
    that $S_{n-1}$ overlaps only $S_n$ and $S_{n-2}$, containing the
    other sets, and that $S_n$ overlaps only $S_{n-1}$ and $S_1$ as it
    contains all other sets in the collection.  Thus, the collection
    $\mathcal{C} = \{S_1, S_2, \ldots, S_n\}$ forms an overlap
    representation for $C_n$ using $n-1$ elements, proving that
    $\varphi(C_n) = n-1$.
  \end{proof}
\end{cor}

We next consider overlap representations of caterpillars.
A tree is a \emph{caterpillar} if the non-leaf vertices
form a path, known as the \emph{spine} of the caterpillar.
We use Theorem~\ref{thm:path}
to find a lower bound on the size of an overlap representation for a
caterpillar, and pair this result with a simple
construction to show that the bound is tight.

\begin{cor}\label{cor:caterpillar}
  For a caterpillar $T$ with spine
  containing $k \geq 1$ vertices, $\varphi(T) = k+2$.

  \begin{proof}
    We show that the size of a minimum overlap representation for
    a caterpillar has size determined by the size of the overlap
    representation for the longest path in the caterpillar.  Let $T$
    be a caterpillar, and label the vertices of the spine in order
    $\{1,2, \ldots, k\}$, and let $L_i$ be the leaves connected to vertex
    $i$ of the spine.  Notice that any longest path in $T$ has a
    vertex in $L_1$ and a vertex in $L_k$ as endpoints, with the
    remaining vertices being those of the spine.  This allows the
    above labelling scheme to be implemented in linear time, as a
    longest path in a tree can be found in linear time.  Also notice
    that the longest path in $T$ contains $k+2$ vertices, and so
    Theorem~\ref{thm:path} provides a lower bound of
    $\varphi(T) \geq k + 2$.

    To show a tight bound, we need only find a representation of the
    correct size.  The representation used is similar to the one
    used in the proof of Theorem~\ref{thm:path}.
    For $T$ a caterpillar, with nodes
    labelled $1,2,\ldots k$ that form the spine, with node $i$
    adjacent to nodes $i-1$ and $i+1$, and $L_i$ the set of leaves
    adjacent to vertex $i$, consider the representation given by, for
    $1 \leq i \leq k$,
    \begin{eqnarray*}
      S_i     & = & \{i+1,i+2\} \\
      S_{L_i} & = & \{1,2, \ldots, i+1\},
    \end{eqnarray*}
    where the set $S_{L_i}$ is associated with all vertices in $L_i$.
    This representation coincides with the one previously given for
    paths, since viewing a path on $n$ vertices as a caterpillar
    produces a caterpillar with $n-2$ vertices on the spine, and two
    leaves, one on each end of the path.  To see that the given
    representation is correct, notice that two vertices of the spine
    $i$ and $j$ overlap if and only if $\abs{i-j}=1$.  Notice also
    that the sets assigned to two leaves never overlap, as
    $S_{L_i}$ overlaps all $S_{L_j}$ for $j \leq i$.  In addition,
    $S_{L_i}$ overlaps only $S_i$, since $S_{L_i}$ contains $S_j$ for
    $j < i$, and $S_{L_i}$ is disjoint from $S_j$ for $j > i$.
    This proves that $\varphi(T) = k+2$.
  \end{proof}
\end{cor}

This representation can be efficiently constructed in the sum of the
sizes of the sets of the representation, 
which is $O(n k)$. 

\subsection{Disconnected Graphs}
\label{scn:components}

In this section we examine the size of a minimum overlap
representation for a disconnected graph based on the sizes of minimum
overlap representation of each of the connected components of the
graph.
This allows us to find the minimum overlap
representation of a graph composed of the pieces we have already
studied, and may lead to divide and conquer algorithms to find the
size of an overlap representation for graphs
such as threshold graphs and cographs that can be defined in terms of
decomposition schemes.

\begin{theorem}\label{thm:components}
  If $G$ is a graph with connected components $B_1,B_2,\ldots,B_k$, then
  \[ \varphi(G) = \sum_{i=1}^k \varphi(B_i) - (k-1). \]

  \begin{proof}
    If $k=1$, the theorem is trivially true.
    We assume that all components of $G$ have size at
    least two, as isolated vertices can be added to a nonempty graph
    without increasing the size of the overlap representation, by
    assigning the isolated vertex a set consisting of any
    single element.  In the case that $G$ consists only of
    isolated vertices, the theorem is also trivially true.  To prove this
    theorem we first, as before, show a lower bound, and then
    argue that a representation achieving this lower bound must exist.

   Suppose $k=2$. By
    Lemma~\ref{lemma:basic-contain}, the two components must either be
    independent, with no elements in common in the overlap
    representation, or some sets of one component can contain all sets
    of the other.  If the two components are independent then
    $\varphi(G) = \varphi(B_1) + \varphi(B_2)$.  In the other case,
    assume without loss of generality that some set associated with a vertex of $B_2$
    contains a set associated with a vertex of $B_1$.
    Thus, by Lemma~\ref{lemma:basic-contain}, any set
    associated with a vertex of $B_2$ that intersects the set $U$ of
    elements in the union of the sets associated with the vertices of
    $B_1$, must contain all of $U$.  In this case the elements of
    $U$ may be considered to act as a single element and so,
    given a minimum overlap representation for $G$, we can take
    the representation restricted to $B_2$ and replace the elements of
    $U$ by a single new element, resulting in an overlap
    representation for $B_2$ of size $\varphi(G) - \varphi(B_1) + 1$.
    Therefore, $\varphi(B_2) \leq \varphi(G) -
    \varphi(B_1) + 1$, which yields the desired bound of

    \[ \varphi(G) \geq \varphi(B_1) + \varphi(B_2) - 1. \]
    
    In the case that $k \geq 3$, we consider a
    minimum overlap representation $\mathcal{C} = \{S_v : v \in V\}$,
    and once again show a lower bound on the size of
    $\mathcal{C}$.  Take any three components with vertex sets $A,
    B,$ and $C$.  If some set associated with a vertex of
    $A$ is contained in a set, $S_b$ for $b \in B$, and some set
    associated with not necessarily the same vertex of $A$ is
    contained in $S_c$ for $c \in C$, then, by
    Lemma~\ref{lemma:basic-contain} the sets $S_b$ and $S_c$ must
    contain $\bigcup_{a \in A} S_a$.  In particular, $S_b$ and $S_c$
    intersect, and so one set must contain the other, as they are sets
    associated with nonadjacent vertices in $G$.  This forces a
    containment relationship between $B$ and $C$, so that the set
    associated with any vertex of $A$ is forced to be contained in
    the sets associated with the vertices of one of $B$ or $C$ by
    transitivity.  To see how this observation is useful, we build a
    graph $F'$, where the vertices of the graph are components in
    $G$, and two vertices $A$ and $B$ are connected by a directed
    edge if there is some vertex $a \in A$ and $b \in B$ such that
    $S_a \subseteq S_b$ in $\mathcal{C}$.  Notice that by
    Lemma~\ref{lemma:basic-contain}, each pair of vertices is
    either nonadjacent, or connected by one directed edge.  The above
    observation is then simply the observation that no vertex, $v$, of
    $F'$ is connected to two nonadjacent vertices by edges
    directed away from $v$.  This implies that if we take the
    transitive reduction of $F'$, we obtain a graph with no cycles,
    and this graph remains acyclic even if we discard the orientation
    of the edges.  Let $F$ be the directed forest resulting from this
    transitive reduction.
    Since the edges of $F$ represent
    containment and no vertex is connected by directed edges to two
    nonadjacent vertices, each tree has a unique root that all edges
    of the tree are directed towards.

    As in the case that $k=2$, if two components
    $B_i$ and $B_j$ are related by containment such that $S_i \subseteq
    S_j$ for some $i \in B_i, j \in B_j$, the elements of $U = \bigcup_{v
    \in B_i} S_v$ function as a single element in the representation for
    the vertices of $B_j$, which is otherwise unrestricted.  Thus if we
    take an overlap representation of these two components we are able
    to find a representation that is at most one element smaller than
    the representation of the two components by disjoint sets.  Notice
    that we can save this one element once for every edge of
    $F$, as these edges count exactly the containment relationships
    that are not forced by transitivity.  The largest number of edges
    $F$ can have is one fewer than the number of components of $G$,
    as there must be some root vertex that is not connected by a
    directed edge to any other vertex.  This provides the following
    lower bound,
    \begin{equation}
      \varphi(G) \geq \sum_{i=1}^k \varphi(B_i) - (k-1).
        \label{eqn:G_minimum}
    \end{equation}

    To show that a representation exists that achieves this bound, we
    take a minimum overlap representation for each component $B_i$ of
    $G$, such that any two of these representations are disjoint.  We
    then, for each $i$ in increasing order, create a containment
    relationship between $B_i$ and $B_{i+1}$, by choosing an arbitrary
    element of the representation for $B_{i+1}$ and replacing it with
    the union of all elements used in the representation of $B_i$.
    The resulting representation is a valid overlap representation for
    $G$, as we have replaced elements in such a way as to not affect
    the overlapping properties within a component, and, given any two
    components, if two sets of the representations associated with
    them have nonempty intersection, then one set must contain the
    other, so that there are no adjacencies created between
    components.  Notice that this representation has size given by
    Equation~\eqref{eqn:G_minimum}, as we have taken optimal
    representations for each component, and removed exactly $k - 1$
    elements, and so this is an optimal overlap representation for
    $G$, of size $\sum_{i=1}^k \varphi(B_i) - (k-1)$, which proves the
    theorem.
  \end{proof}
\end{theorem}

\section{Hardness Results}
\label{scn:hardness}

In this section we present some \class{NP}-completeness results
for problems related to finding the minimum overlap representation of
a given graph.  

\subsection{Extending a Representation}
\label{scn:extension-hard}

A natural approach to finding the overlap number for a graph is to
employ a greedy strategy, adding one vertex at a time, and only making
changes to the set associated with the newly added vertex.
Unfortunately, this is not a feasible approach for a general graph and
overlap representation, as
the problem of deciding whether or not a new element needs to be added
to the representation is \class{NP}-complete.  The formal statement of
this decision problem is as follows.

\begin{problem}
    The \prob{Overlap Extension} problem is defined as:
    \begin{description}
        \item[Instance:] A graph, $G = (V, E)$, an overlap representation
          $\mathcal{C} = \{S_v : v \in V\}$ of $G$, and a set $A
          \subseteq V$.

        \item[Question:] Is
	      there a set $S \subseteq \bigcup_{v \in V} S_v$
	      that overlaps $S_v$ if and only if $v \in A$?
    \end{description}
\end{problem}

Since such an extension can be efficiently verified, this problem is
in $\class{NP}$.  To see that the related problem on intersection
representations can be solved efficiently, notice that in the
intersection case, an element $i$ of the representation can be added
to $S$ if and only if the set $A$ contains all vertices $v$ with $i
\in S_v$.  If all elements that can be added to $S$ fail to form an
intersection representation for the extended graph, then without
introducing a new element, no such extension is possible.

Returning to the overlap case, the problem that we reduce to
\prob{Overlap Extension} is the \prob{Not-All-Equal 3SAT} problem,
which is identical to the standard \prob{3SAT} problem, with the
exception that we seek a satisfying truth assignment where no clause
has all true literals.  This problem is known to be
\class{NP}-complete~\cite{schaefer78complexity}.

\begin{theorem} \label{thm:extension-hard}
  \prob{Overlap Extension} is \class{NP}-complete.

  \begin{proof}
    Let $(U,F)$ be an instance of
    \prob{Not-All-Equal 3SAT}, where $U = \{x_1, x_2, \ldots, x_n\}$
    is the set of variables, and $F = \{c_1, c_2, \ldots, c_m\}$ is
    the set of clauses with $\abs{c_i} = 3$, for each $i$.
    If $n < 4$, we can examine all possible truth assignments to
    determine if there is a solution to the \prob{Not-All-Equal 3SAT}
    instance, and output a trivial yes or no instance of \prob{Overlap
    Extension}.

    If $n \geq 4$, we construct a graph $G =
    (V,E)$, an overlap representation $\mathcal{C}$ of $G$, and a set
    $A$, to form an instance of \prob{Overlap Extension}.  The
    vertices in the graph are given by
    \[
    V = \{v_i : 1 \leq i \leq n\} 
        \cup \{w_i : 1 \leq i \leq m\},
    \]
    where each $v_i$ is associated with a variable $x_i \in U$,
    and each $w_i$ is associated with a clause $c_i \in
    F$.  We take the overlap representation
    representation $\mathcal{C}$ given by
    \[
      \mathcal{C}
      = \{S_{v_i} = \{x_i, \neg x_i\} : x_i \in U\}
      \cup \{S_{w_i} = c_i : c_i \in F\},
    \]
    and we set $E$ to those edges consistent this representation.
    Finally, we let $A = V$ to complete the instance
    $(G,\mathcal{C},A)$ of \prob{Overlap Extension}.  This
    transformation can clearly be performed in polynomial time.  A
    solution of the extension problem is a set of literals that
    overlaps each set in $\mathcal{C}$, and we show that such a
    set is equivalent to a satisfying truth assignment for $(U,F)$ in
    which each clause has at least one false literal.

    To see this, let $S \subseteq U \cup \{\neg x : x \in U\}$ be a
    set that overlaps all elements of $\mathcal{C}$.  Since $S$
    overlaps each $S_{v_i} = \{x_i, \neg x_i\}$, $S$ must contain
    exactly one element of $S_{v_i}$, and so we consider the truth
    assignment $T$ that makes each literal in $S$ true.
    In addition, $S$
    overlaps each $S_{w_i} = c_i$, which forces at least one, but not
    all, of the literals in $c_i$ to be contained in $S$, which shows
    that $T$ satisfies the clause $c_i$ without making all literals true.

    In the other direction, we take any truth assignment $T$ that
    satisfies $(U,F)$ without making all literals in any clause true,
    and consider the set $S$ of all literals made true by $T$.  
    Since $T$ is a truth
    assignment, for each $1 \leq i \leq n$, $S$ contains exactly one
    of $x_i$ and $\neg x_i$, and so $S$ overlaps $S_{v_i}$ for all
    $i$.  Furthermore, since $T$ is a satisfying truth assignment, $S$
    must intersect each $S_{w_i}$, and it cannot contain any
    $S_{w_i}$, as this would imply that $T$ satisfies all literals
    of each clause $c_i$.  Finally, $\abs{S_{w_i}} = \abs{c_i} = 3$, and
    $\abs{S} = \abs{U} \geq 4$, so $S_{w_i}$ cannot contain $S$ for
    any $i$.  This implies that $S$ overlaps $S_{w_i}$ for all $1 \leq
    i \leq m$, and so $S$ is a solution to the instance of
    \prob{Overlap Extension}.
  \end{proof}
\end{theorem}

Using a similar reduction, we can show the hardness of the problem of
the \prob{Containment Extension} problem, which is the analogue of the
\prob{Overlap Extension} problem on containment representations.  In
this case the reduction is from the well-known \class{NP}-complete
\prob{3SAT} problem.

\begin{theorem}\label{thm:containment-extension-hard}
  \prob{Containment Extension} is \class{NP}-complete.

  \begin{proof}
    Let $(U,F)$ be an instance of \prob{3SAT}, where $U = \{x_1,
    x_2, \ldots, x_n\}$ is a set of $n$ variables, and $F = \{c_1, c_2,
    \ldots, c_m\}$ is a set of $m$ clauses, each containing three
    literals.  We may once again consider only the case where $n \geq 4$,
    as the reduction can output a trivial yes or no instance if this
    is not the case.

    The vertices of the constructed graph $G = (V,E)$ are given, similarly
    to the \prob{Overlap Extension} case, by
    \[ V = \{ v_i : 1 \leq i \leq n\} 
           \cup \{w_i : 1 \leq i \leq m\} \cup \{z\}. \]
    We set $L = \bigcup_{i=1}^n \{x_i, \neg x_i\}$, the set of all
    literals, and construct the containment representation given by the
    collection $\mathcal{C}$ consisting of the following sets, for all
    $1 \leq i \leq n$ and $1 \leq j \leq m$,
    \begin{eqnarray*}
       S_{v_i} & = &  \{x_i, \neg x_i\} \\
       S_{w_i} & = & (L \setminus c_i) \cup \{0\} \\
       S_z & = & \{0\}.
    \end{eqnarray*}
    To complete the constructed instance, we set $A = \{z\}$.

    In a similar way to the proof of Theorem~\ref{thm:extension-hard},
    it can be observed that there is a set extending this containment
    representation if and only if the original instance of \prob{3SAT}
    has a satisfying truth assignment.  The idea is that any such set
    $S$ must contain the element $0$, and so $S$ cannot be contained
    in $L \setminus c_i$ for any clause $c_i$, which is exactly the
    requirement that $S$ contains a literal in $c_i$.  In addition,
    the truth assignment given by $S$ must form a valid partial truth
    assignment, since $S$ can contain at most one of each pair of
    literals, as it cannot contain any set $S_{v_i}$.  The other
    direction is again similar to the proof of
    Theorem~\ref{thm:extension-hard}, as the set of all literals a
    satisfying truth assignment makes true is a valid extension
    of the containment representation.
  \end{proof}
\end{theorem}

\subsection{Containment-Free Representations}
\label{scn:cf-hard}

In the remainder of this section, we consider the problem of finding a
minimum overlap representation where no set is contained in any other,
or where the number of set containments is limited.

\begin{problem}
    The \prob{CF-Overlap Number} problem is defined as:
    \begin{description}
        \item[Instance:] A graph, $G = (V, E)$, and a natural number $k$.

        \item[Question:] Does
          the graph $G$ have a containment-free overlap representation
          of size $k$?
    \end{description}
\end{problem}

In the absence of containment, the definitions of overlap and
intersection coincide, and so this problem is equivalent to the
problem of finding a minimum containment-free intersection
representation.  In order to show the hardness of this problem, we
reduce the \prob{Intersection Number} problem to it, since
\prob{Intersection Number} is known to be
\class{NP}-complete~\cite{ksw78covering}.

\begin{theorem}\label{thm:containment-free}
  \prob{CF-Overlap Number} is \class{NP}-complete.
  
  \begin{proof}
    Given an instance
    $G = (V,E)$ and $k$ of \prob{Intersection Number}, with
    $n = \abs{V}$, we construct the graph $G'$ by adding, for each $v
    \in V$, a new vertex $v'$ that is adjacent only to $v$.  Let
    $V'$ be the set of all new vertices in $G'$, and let $E'$ be the
    set of all edges in $G'$ incident on a vertex in $V'$.
    The instance of \prob{CF-Overlap Number} is then given by
    $G'$ and $k+2n$, which can clearly be constructed in polynomial time.

    Notice that any
    containment-free overlap representation forms a containment free
    intersection representation, and further that in any
    containment-free intersection representation for $G'$, the sets
    $S_v$ and $S_{v'}$ associated with a vertex $v \in V$ and $v' \in
    V'$ must share a common element, as these vertices are adjacent,
    and furthermore, since $v'$ is adjacent only to $v$,
    this element is only be found in $S_v$ and $S_{v'}$.  The set
    $S_{v'}$ is not contained in $S_v$, and so it must contain at
    least one other element, which is unique to the set
    $S_{v'}$, since $v'$ is adjacent only to $v$.
    This implies that for all $v \in V$, there are at least two
    elements found only in one or both of $S_v$ and $S_{v'}$, which
    ensures that there are no containment relationships between any
    sets of the representation.
    Since these elements suffice to represent the vertices in $V'$,
    and the representation is already containment free, the
    remaining elements of the representation form an arbitrary
    intersection representation for $G$.  Hence, the containment-free
    overlap number of $G'$ is exactly $\theta_e(G) + 2n$,
    where $\theta_e(G)$ is the size of a minimum intersection
    representation for $G$.  Thus $G$ has an intersection
    representation of size $k$ if and only if $G'$ has a
    containment-free overlap representation of size $k + 2n$.
  \end{proof}
\end{theorem}

\subsection{Overlap Representations with Limited Containment}
\label{scn:hardness-lim-cont}

We can extend the hardness of the \prob{CF-Overlap Number} problem
to the problem of finding a minimum overlap representation of a graph,
using at most a constant number of containment relationships between
sets of the representation.
Formalized as a decision problem, we consider the following problem.
The factor of $2$ appears since the nonadjacent pairs $(u,v)$ and $(v,u)$
are both counted, but we refer to the single non-edge as a containment
relationship.

\begin{problem}
    The \prob{$L$-Containment Overlap Number} problem, for any
    natural number $L$, is defined as:
    \begin{description}
        \item[Instance:] A graph, $G = (V, E)$, and a natural number $k$.
        \item[Question:] Is
	      there is some collection $\mathcal{C} = \{S_v : v \in V\}$
          that forms an overlap representation, such that
          $\abs{\bigcup_{v \in V} S_v} \leq k$ and
          $ \abs{\{(u,v) \not\in E : u \neq v \text{ and }
            S_u \cap S_v \neq \emptyset\}}
          \leq 2L$?
    \end{description}
\end{problem}

This problem, when $L = 0$ is exactly the \prob{CF-Overlap Number}
problem and so by Theorem~\ref{thm:containment-free} it is
\class{NP}-complete in this case.
For any constant $L$, a simple
Turing reduction from the \prob{CF-Overlap Number} problem is
given by making $2L + 1$ copies of the input graph, and then finding
an overlap representation with no more than $L$ containments, which, by
the pigeonhole principle, must leave at least one copy of $G$
containment free, both internally, and with respect to other
components of the graph.  Furthermore, if we have a minimum representation,
then this representation for $G$ must also be minimum, as the sets
associated with the vertices of this copy of $G$ are disjoint from the
sets associated with vertices in any other copy.
With a little more work, we can find a many-one reduction from the
\prob{CF-Overlap Number} problem, by adding to the graph $G$ extra
components where a minimum representation is be compelled to
``spend'' all $L$ set containments, leaving $G$ with a
containment-free representation.  We can do this in such a way that we
can track the number of elements these extra components add to
the representation.
To show this result we make use of
Corollary~\ref{cor:contain-connected}, which gives an upper
bound on the number of elements we are able to save by allowing
containment relationships between components of the constructed graph.

\begin{theorem}
  For any $L \in \mathbb{N}$,
  \prob{$L$-Containment Overlap Number} is \class{NP}-complete.
  
  \begin{proof}
    Let $G = (V,E)$ and $k$ be an instance of \prob{CF-Overlap Number}.
    We set $n = \abs{V}$, and we consider only cases where $n \geq 4$, as
    smaller cases can be solved as part of the transformation by
    searching all possible representations and producing as output a
    trivial yes or no instance.
    In the instance we construct, we add $2L$
    components to the graph $G$.  Each of these components is given by
    the graph $B_i = (V_i, E_i)$, which is constructed from $n+1$
    disjoint edges, with three nonadjacent universal vertices, as shown
    in Figure~\ref{fig:loverlap_reduction}.  
    More formally, the vertices $V_i$ and the edges $E_i$ of each
    component $B_i$ are given by
    \begin{eqnarray*}
      V_i & = & \{v_{i,j} : 1 \leq j \leq 2n+2\} \cup \{x_i,y_i,z_i\},\\
      E_i & = & \{(v_{i,2j-1}, v_{i,2j}) : 1 \leq j \leq n + 1\} \cup \\
          &   & \{(x_i, v_{i,j}), (y_i,v_{i,j}), (z_i,v_{i,j}) 
                  : 1 \leq j \leq 2n + 2\}.
    \end{eqnarray*}
\begin{figure}
  \begin{center}
    \ifpdf
      \input{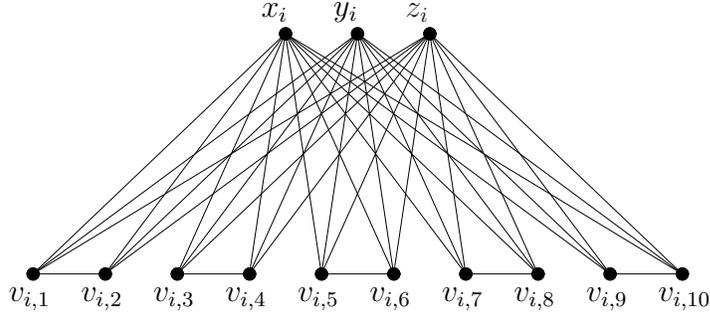}
    \else
      \setlength{\unitlength}{0.00083333in}
\begingroup\makeatletter\ifx\SetFigFont\undefined%
\gdef\SetFigFont#1#2#3#4#5{%
  \reset@font\fontsize{#1}{#2pt}%
  \fontfamily{#3}\fontseries{#4}\fontshape{#5}%
  \selectfont}%
\fi\endgroup%
{\renewcommand{\dashlinestretch}{30}
\begin{picture}(4880,2029)(0,-10)
\put(2475,1858){\makebox(0,0)[lb]{{\SetFigFont{12}{14.4}{\familydefault}{\mddefault}{\updefault}$z_i$}}}
\put(600,283){\blacken\ellipse{76}{76}}
\put(600,283){\ellipse{76}{76}}
\path(150,283)(600,283)
\put(1050,283){\blacken\ellipse{76}{76}}
\put(1050,283){\ellipse{76}{76}}
\put(1500,283){\blacken\ellipse{76}{76}}
\put(1500,283){\ellipse{76}{76}}
\path(1050,283)(1500,283)
\put(3750,283){\blacken\ellipse{76}{76}}
\put(3750,283){\ellipse{76}{76}}
\put(4200,283){\blacken\ellipse{76}{76}}
\put(4200,283){\ellipse{76}{76}}
\path(3750,283)(4200,283)
\put(1950,283){\blacken\ellipse{76}{76}}
\put(1950,283){\ellipse{76}{76}}
\put(2400,283){\blacken\ellipse{76}{76}}
\put(2400,283){\ellipse{76}{76}}
\path(1950,283)(2400,283)
\put(1725,1783){\blacken\ellipse{76}{76}}
\put(1725,1783){\ellipse{76}{76}}
\put(2175,1783){\blacken\ellipse{76}{76}}
\put(2175,1783){\ellipse{76}{76}}
\put(2625,1783){\blacken\ellipse{76}{76}}
\put(2625,1783){\ellipse{76}{76}}
\put(2850,283){\blacken\ellipse{76}{76}}
\put(2850,283){\ellipse{76}{76}}
\put(3300,283){\blacken\ellipse{76}{76}}
\put(3300,283){\ellipse{76}{76}}
\path(2850,283)(3300,283)
\path(150,283)(1725,1783)
\path(1725,1783)(600,283)
\path(1050,283)(1725,1783)
\path(1725,1783)(1500,283)
\path(1950,283)(1725,1783)
\path(2400,283)(1725,1783)
\path(1725,1783)(2850,283)
\path(3300,283)(1725,1783)
\path(1725,1783)(3750,283)
\path(2175,1783)(150,283)
\path(600,283)(2175,1783)
\path(2175,1783)(1050,283)
\path(1500,283)(2175,1783)
\path(2175,1783)(1950,283)
\path(2175,1783)(2400,283)
\path(2175,1783)(2850,283)
\path(2175,1783)(3300,283)
\path(2175,1783)(3750,283)
\path(4200,283)(2175,1783)
\path(2625,1783)(150,283)
\path(600,283)(2625,1783)
\path(2625,1783)(1050,283)
\path(1500,283)(2625,1783)
\path(2400,283)(2625,1783)
\path(2625,1783)(2850,283)
\path(3300,283)(2625,1783)
\path(2625,1783)(3750,283)
\path(4200,283)(2625,1783)
\path(4200,283)(1725,1783)
\path(2625,1783)(1950,283)
\put(0,58){\makebox(0,0)[lb]{{\SetFigFont{12}{14.4}{\familydefault}{\mddefault}{\updefault}$v_{i,1}$}}}
\put(450,58){\makebox(0,0)[lb]{{\SetFigFont{12}{14.4}{\familydefault}{\mddefault}{\updefault}$v_{i,2}$}}}
\put(900,58){\makebox(0,0)[lb]{{\SetFigFont{12}{14.4}{\familydefault}{\mddefault}{\updefault}$v_{i,3}$}}}
\put(1350,58){\makebox(0,0)[lb]{{\SetFigFont{12}{14.4}{\familydefault}{\mddefault}{\updefault}$v_{i,4}$}}}
\put(2250,58){\makebox(0,0)[lb]{{\SetFigFont{12}{14.4}{\familydefault}{\mddefault}{\updefault}$v_{i,6}$}}}
\put(2700,58){\makebox(0,0)[lb]{{\SetFigFont{12}{14.4}{\familydefault}{\mddefault}{\updefault}$v_{i,7}$}}}
\put(3150,58){\makebox(0,0)[lb]{{\SetFigFont{12}{14.4}{\familydefault}{\mddefault}{\updefault}$v_{i,8}$}}}
\put(3600,58){\makebox(0,0)[lb]{{\SetFigFont{12}{14.4}{\familydefault}{\mddefault}{\updefault}$v_{i,9}$}}}
\put(4050,58){\makebox(0,0)[lb]{{\SetFigFont{12}{14.4}{\familydefault}{\mddefault}{\updefault}$v_{i,10}$}}}
\put(1800,58){\makebox(0,0)[lb]{{\SetFigFont{12}{14.4}{\familydefault}{\mddefault}{\updefault}$v_{i,5}$}}}
\put(1575,1858){\makebox(0,0)[lb]{{\SetFigFont{12}{14.4}{\familydefault}{\mddefault}{\updefault}$x_i$}}}
\put(2025,1858){\makebox(0,0)[lb]{{\SetFigFont{12}{14.4}{\familydefault}{\mddefault}{\updefault}$y_i$}}}
\put(150,283){\blacken\ellipse{76}{76}}
\put(150,283){\ellipse{76}{76}}
\end{picture}
}
    \fi
  \end{center}
  \caption[Example of a component in the reduction]{Example of $B_i$
    with $n = 4$.}\label{fig:loverlap_reduction}
\end{figure}
    The graph in the constructed instance of
    \prob{$L$-Containment Overlap Number} is then given by a
    disjoint union, $H = G + B_1 + B_2 + \cdots + B_{2L}$, of $2L$ of
    these new components with the graph $G$.
    The value $k'$ is set to
    \begin{equation}\label{eqn:reduction-cf-l-bound}
      k' = k + 3L(n+1) + 4L(n+1),
    \end{equation}
    to complete the instance of \prob{$L$-Containment Overlap Number}.

    Before showing that the given instance of the \prob{CF-Overlap
      Number} problem is equivalent to the constructed
    instance, we first make some observations about overlap
    representations of the graphs $B_i$.  In a minimum containment-free
    overlap representation for the vertices $v_{i,j}$ of $B_i$, we
    must use $3(n+1)$ elements, as each disjoint edge $(v_{i,2j-1},
    v_{i,2j})$ requires at least three new elements in the
    representation.
    Furthermore, these three elements are given by an element unique to
    $S_{v_{i,2j-1}}$, an element unique to $S_{v_{i,2j}}$, and an
    element in the intersection of these two sets.
    We can extend a minimum representation for these vertices to include
    $x_i$ and $y_i$ without increasing the size of the
    representation.  
    To do this, we set $S_{x_i}$ to be those elements in common to the
    sets associated with both endpoints of each edge $(v_{i,2j-1},
    v_{i,2j})$, and we set
    $S_{y_i}$ to the elements that are unique to each of these sets.
    Since $n + 1 \geq 2$, these sets are
    not be contained in any set $S_{v_{i,j}}$, and so these sets
    overlap, as desired.  
    This forms the unique (up to permutation of the elements) minimum
    containment-free representation for all the vertices of $B_i$
    except $z_i$.  If we allow a single containment relationship, we
    can set $S_{z_i} = S_{x_i}$ to obtain a representation with size
    $3(n+1)$.

    If we seek a containment-free overlap representation for $B_i$ the
    situation is more bleak, as we cannot extend the unique minimum
    containment representation for every vertex except $z_i$ without
    adding new elements.  
    This is because we still must use three elements to represent each
    edge $(v_{i,2j-1},v_{i,2j})$, but there is no partition of these
    elements into three sets such that each set overlaps both
    $S_{v_{i,2j-1}}$ and $S_{v_{i,2j}}$.
    We are required then to use four elements for each edge
    $(v_{i,2j-1},v_{i,2j})$, with two elements in common to the sets
    associated with the endpoints, which brings the size of a minimum
    containment-free overlap representation of $B_i$ to $4(n+1)$.
    The key to the remainder of the proof is that by allowing a single
    containment relationship, we can reduce the size of the
    representation for some component $B_i$ by $n+1$ elements.

    If $\mathcal{C}$ is a minimum $L$-overlap representation for $H$,
    of size no more than $k' = k + 3L(n+1) + 4L(n+1)$, we show
    that $G$ has a containment-free overlap representation of size not
    more than $k$.  We claim that, as $\mathcal{C}$ is minimum, the
    representation $\mathcal{C}$ when restricted to $G$ is already
    containment-free, and in fact, the $L$ containment
    relationships can be found in $L$ of the components $B_i$.
    To show this, we examine the other potential
    cases for a non-edge to be represented by containment, showing in
    each one that we can make a local transformation to move the
    containment relationship to some component $B_i$, in the process
    reducing the size of the representation, contradicting the
    optimality of $\mathcal{C}$.

    The first such case we consider is any containment within the
    representation of $G$, which is, two vertices $u$ and $v$ such
    that $S_u \subseteq S_v$.  We replace $S_v$ with $n-1$ new
    elements, $a_1, a_2, \ldots, a_{n-1}$, to obtain $S'_v = \{a_1,
    a_2, \ldots, a_{n-1}\}$.  This removes the containment
    relationship between $u$ and $v$, and forces $S'_v$ not to contain
    any other set in the representation.  In order to ensure that the
    representation is still valid,
    we modify the sets associated with some of the other
    vertices.  There are exactly three ways a set can interact with
    $S'_v$: we can have the set we consider contain $S'_v$, the two
    sets can be disjoint, or the two sets can overlap.
    We consider, for each of these three interactions, how to
    alter the set to maintain a valid overlap representation of $H$.
    For any vertex $w$ with $S_v \subseteq S_w$, we replace the set
    $S_w$ with the set $S'_w = S_w \cup S'_v$, to ensure that this
    containment relationship is not altered.  This alteration does not
    affect the overlap, containment, or disjointedness relationships of
    the set $S_w$, as these are new elements, and by transitivity, we
    have added these new elements to any set that contains $S_w$.  If
    $w$ is a vertex such that $S_w$ and $S_v$ are disjoint, then $S_w$
    and $S'_v$ must also be disjoint, and there is nothing to do in
    this case.  If $w$ is such that $S_w$ and $S_v$ overlap, then we
    have $S'_v \cap S_w = \emptyset$, which we correct by setting
    $S'_w = S_w \cup \{a_i\}$, for an element $a_i \in S'_v$ that we
    have not already used for this purpose.  This forces $S'_w$
    and $S'_v$ to overlap, as the conditions that $n \geq 4$ and $S_w$
    overlaps the set $S_v$ ensure that there are least two elements
    in each of these sets.  We must also add the element $a_i$ to any
    set that contains $S_w$ to preserve this containment
    relationship.  This does not affect the representation of any
    vertex but $v$, as the sets that the element $a_i$ is being added
    to must also intersect $S_v$, and we do not
    add all of the $a_i$ to a set that should not contain $S'_v$,
    since there are at most $n-2$ vertices that are adjacent to $v$.
    Thus, we can remove at least one containment
    relationship from $G$, by adding $n-1$ new elements to the
    representation.  Since there are $2L$ components $B_i$, and only
    $L-1$ remaining containments, 
    there must be some $i$ for which the vertices of
    $B_i$ are involved in no containment relationships.  We can use
    the containment we just removed from $G$ to reduce the size of the
    representation for $B_i$ from $4(n+1)$ to $3(n+1)$, which, in
    total, saves at least $n + 1 - (n-1) = 2$ elements from the
    representation, contradicting the assumption that $\mathcal{C}$
    was minimal.  Thus, the vertices of $G$ are not involved in any
    containment relationships in $\mathcal{C}$.

    The second case of a containment relationship is one internal to
    one of the components $B_i$.  If this containment is between two
    vertices $v_{i,j}$ and $v_{i,k}$, we can simply replace the
    representation for $v_{i,j}$ and the vertex $v_{i,j \pm 1}$ it forms
    an edge with.
    This is done by setting $S_{v_{i,j}} = \{a_1, a_3,
    a_4\}$ and $S_{v_{i,j \pm 1}} = \{a_2, a_3, a_4\}$, where the elements
    $a_i$ are new to the representation.  Finally, we add $a_1$ and
    $a_2$ to $S_{x_i}$, $a_3$ to $S_{y_i}$, and $a_4$ to $S_{z_i}$,
    being careful to add these elements to any set that contains these
    elements.
    If preserving these containment relationships results in all of
    $\{a_1,a_2,a_3,a_4\}$ being contained in one of $S_{x_i},
    S_{y_1},$ or $S_{z_i}$,
    we simply add $a_3$ to each of $S_{x_i}, S_{y_i}$ and $S_{z_i}$, and
    remove $a_1,a_2,$ and $a_4$ from these sets, once again being
    careful to preserve any containment relationships.
    This replacement removes the containment between $v_{i,j}$ and
    $v_{i,k}$, and leaves a valid overlap representation.  As the cost
    of this alteration was only four elements, and we can apply the
    freed containment relationship to some other component $B_j$ to
    save $n+1$ containments, this also contradicts the optimality
    of $\mathcal{C}$.

    The only remaining case for a
    containment internal to $B_i$ is one between two of $x_i, y_i$,
    and $z_i$, as these vertices are adjacent to all other vertices of
    $B_i$.  We must also consider the case that between the vertices
    $x_i, y_i$ and $z_i$ there are two or more containments, but since
    we can extend a minimum representation for the vertices $v_{i,j}$
    to these vertices using only one containment, we can again apply
    this containment elsewhere, contradicting the optimality of $\mathcal{C}$.

    The final case we must consider is a containment relationship
    between two vertices in differing components of $H$.  Let $U$ and
    $W$ be the vertex sets of the two components, where for some
    vertex $u \in U$ and $w \in W$ we have $S_u \subseteq S_w$.
    Let $A = \bigcup_{u \in U} S_u$.  Lemma~\ref{lemma:basic-contain}
    implies that for any vertex in $v \in W$, either $S_v$ contains
    $A$ or it is disjoint from it.
    The elements of $A$ then, within $W$, act as a single element.  This
    allows these elements to be replaced with a single new element,
    where once again, whenever we add a new element to a set we must
    also add this new element to any sets that contained
    the original set.  After this replacement has been made, we have
    removed at least one containment relationship, at a cost of one
    new element in the representation, which once again contradicts
    the optimality of $\mathcal{C}$.
    
    Thus, a minimum $L$-containment overlap representation for $H$
    uses containment only between the vertices $x_i, y_i,$ and $z_i$,
    and uses at most one containment per triple of vertices.  Thus,
    in a minimum overlap representation, we have a containment-free
    overlap representation for $G$, and $L$ of the $B_i$,
    and we have an overlap representation using only one
    containment for the remaining $L$ of the $B_i$.  Then,
    where we $r$ is the containment-free overlap number
    of $G$, this representation has size $r + 4L(n+1) + 3L(n+1)$,
    which by Equation~\eqref{eqn:reduction-cf-l-bound} is less than
    $k'$ only when $r \leq k$, as desired.

    Fortunately, the other direction is simple.  If we take any
    containment-free overlap representation for $G$ of size no more
    than $k$, we can form the representations discussed above for each
    $B_i$, by simply using three elements per edge $(v_{i,2j-1},
    v_{i,2j})$ for $L$ of the $B_i$ and four elements per edge for the
    remaining $L$.  Placing the containments in appropriate places, we
    can find an $L$-containment overlap representation for $H$ of size
    no more than $k + 3L(n+1) + 4L(n+1) = k'$, as required.
  \end{proof}
\end{theorem}

\section{Conclusion}

There are many open problems related to the overlap number of a
graph.
Foremost among these unanswered questions
is the complexity of computing the overlap number.

\begin{problem}
    The \prob{Overlap Number} problem is defined as:
    \begin{description}
        \item[Instance:] A graph, $G = (V, E)$, and an integer $k$.

        \item[Question:] Is
	      there an overlap representation $\mathcal{C} =
	      \{S_v : v \in V\}$ of $G$ with
          $\abs{ \bigcup_{v \in V} S_v } \leq k$?
    \end{description}
\end{problem}

This problem is clearly in \class{NP}, as it is a simple matter to
verify that a given representation is both correct and of the
appropriate size, and the evidence suggests that this problem is
also complete for \class{NP}.
There are also many class of graphs for
which no algorithm to find a minimum overlap representation is known.
Many of these classes, such as cographs, are classes
of graphs for which many other combinatorial problems are tractable,
and so there is reason to believe that efficient algorithms exist to
compute the overlap number on some such classes of graphs, but they have
yet to be discovered.

\section*{Acknowledgements}

This research was partially supported by {NSERC} and {iCORE}.

\bibliographystyle{abbrv}
\bibliography{overlap}

\newpage
\appendix
\section*{\Large\bf \centerline{Appendix}}
\setcounter{section}{1}

This appendix contains the proofs that have been 
omitted from the main text.

\subsection*{Proofs Omitted From Section~\ref{scn:algo}}

\begin{namedtheorem}{Lemma \ref{lemma:messy_stirling}}
  For $1 \leq k < n$,
  \[ \binom{n}{k} \geq \sqrt{\frac{1}{8 \pi k}}
      \left(\frac{n}{k}\right)^k \left(\frac{n}{n-k}\right)^{n-k}. \]
  \begin{proof}
    By simple expansion, using Stirling's Approximation
    (Equation~\eqref{eqn:stirling}), we have
    \begin{eqnarray*}
      \binom{n}{k} = \frac{n!}{k! (n-k)!}
      & \geq & \frac{\sqrt{2 \pi n} (n/e)^n}
                    {2 \pi e^{1/(12k) + 1/(12(n-k))}
		      \sqrt{k(n-k)} (k/e)^k ((n-k)/e)^{n-k}} \\
      & \geq &  \frac{1}{e^{1/6}} \sqrt{\frac{n}{2 \pi k(n-k)}}
               \frac{n^n}{k^k (n-k)^(n-k)} \\
      & \geq &  \frac{1}{2} \sqrt{\frac{n}{2 \pi k(n-k)}}
               \left(\frac{n}{k}\right)^k \left(\frac{n}{n-k}\right)^{n-k} \\
      & \geq &  \sqrt{\frac{n-k}{8 \pi k(n-k)}}
               \left(\frac{n}{k}\right)^k \left(\frac{n}{n-k}\right)^{n-k} \\
      & = &  \sqrt{\frac{1}{8 \pi k}}
               \left(\frac{n}{k}\right)^k \left(\frac{n}{n-k}\right)^{n-k}
    \end{eqnarray*}
    as in the statement of the lemma.
  \end{proof}
\end{namedtheorem}

\begin{namedtheorem}{Theorem \ref{thm:clique-bound}}
  For $n \ge 1$,
  \[ \min \left\{m : n 
    \leq \binom{m}{\floor{\frac{m+2}{2}}} \right\} \in \Theta(\log n).\] 

  \begin{proof}
    Let 
     $ x = \min \left\{m : n 
        \leq \binom{m}{\floor{\frac{m+2}{2}}} \right\}$.
    We show a lower bound by observing that there are
    $2^x$ subsets of $\{1,2, \ldots, x\}$, and so we must
    have $n < 2^x$, which implies that $x \in \Omega(\log n)$.
    Turning to an upper bound, notice that, by
    the definition of $x$, we have
    $ \binom{x-1}{\floor{\frac{x+1}{2}}} <  n. $
    Using this, and Lemma~\ref{lemma:messy_stirling}, we have
    \begin{eqnarray*}
      n 
      & > & \binom{x-1}{\floor{\frac{x+1}{2}}}
       \geq  \binom{x-1}{\frac{x+1}{2}}
       \geq  \sqrt{\frac{2}{8 \pi (x+1)}}
               \left(\frac{2(x-1)}{x+1}\right)^{(x+1)/2} 
               \left(\frac{2(x-1)}{x-1}\right)^{(x-1)/2} \\
      & = &    2^x \sqrt{\frac{1}{4 \pi (x+1)}}
               \left(\frac{x-1}{x+1}\right)^{(x+1)/2}
       \geq  2^{(x-1)/2} \sqrt{\frac{1}{4 \pi (x+1)}} \\             
    \end{eqnarray*}
    We can then take logarithms to obtain
    \begin{eqnarray*}
      \log n 
      & > & \log \left( 2^{(x-1)/2} \sqrt{\frac{1}{4 \pi (x+1)}}\right)
       >  \frac{x-1}{2} + 
            \frac{1}{2} \log \left( \frac{1}{4 \pi (x+1)} \right)\\
      & = & \frac{x-1}{2} -
            \frac{\log \left(4 \pi (x+1)\right)}{2}
       =  \frac{x-1}{2} -
            \frac{\log 4 \pi + \log (x+1)}{2}\\
    \end{eqnarray*}
    Since $x \in \Omega(\log n)$, for large enough $n$ we must have
    $\log{(x+1)} < x/2$.  We then have, by
    the above, and setting $C = \log(4\pi)/2$,
    \[
      \log n > \frac{x-1}{2} - \frac{x}{4} - C = \frac{x-2}{4} - C,
    \]
    which is $x/4 < \log n + 1/2 + C$,
    and so we have $x \in O(\log n)$, as desired.
  \end{proof}
\end{namedtheorem}

\end{document}